# Unstable Angina is a syndrome correlated to mixed Th17 and Th1 immune disorder

By Wan-Chung Hu*
*Postdoctorate
Genomics Research Center
Academia Sinica
No 128 Academia Road section2
Nangang 115, Taipei, Taiwan

**Previous Institute:**

Department of Neurology
Shin Kong Memorial Hospital
Taipei, Taiwan

**Abstract**

Unstable angina is common clinical manifestation of atherosclerosis. However, the detailed pathogenesis of unstable angina is still not known. Here, I propose that unstable angina is a mixed TH17 and TH1 immune disorder. By using microarray analysis, I find out that TH1 and TH17 related cytokine, cytokine receptor, chemokines, complement, immune-related transcription factors, anti-bacterial genes, Toll-like receptors, and heat shock proteins are all up-regulated in peripheral leukocytes of unstable angina. In addition, H-ATPase, glycolytic genes, platelet and RBC related genes are also up-regulated in peripheral leukocytes of during unstable angina. It also implies that atherosclerosis is a mixed TH17 and TH1 autoimmune disease. If we know the etiology of unstable angina as well as atherosclerosis better, we can have better methods to control and prevent this detrimental illness.

**Introduction**

Atherosclerosis is a very popular disease, especially in developed countries. It is the top death-causing disease in industrial countries. Coronary artery disease (CAD) is the most important clinical symptom of atherosclerosis. Typical angina and unstable angina are common manifestations of CAD. Atherosclerosis is thought to be an inflammatory disorder. However, the detailed pathogenesis of angina is not known. Here, I use microarray analysis of peripheral leukocytes to describe unstable angina is a mixed TH1 and TH17 immune disorder.

**Materials and methods**

Study samples

I use microarray dataset from Gene Omnibus website (GEO). The dataset serial number is GSE. This dataset includes peripheral leukocytes from patients of acute myocardial infarction and unstable angina. Here, I compare the RNA expression profile of leukocytes from unstable angina patients to that from health normal control. This dataset is GSE29111 from Dr. Milo and Rothman's study group. The healthy normal controls are from dataset of a Huntington's disease blood biomarker study population. The dataset is GSE8762 from Dr. Runne H.'s study group. This second study was published in PNAS,2007, 104(36):14424.

RMA normalization

Affymetrix HG-U133A 2.0 genechip was used in both samples. RMA express software(UC Berkeley, Board Institute) is used to do normalization and to rule out the outliners of the above dataset. I rule out the potential outliners of samples due to the following criteria:

1. Remove samples which have strong deviation in NUSE plot
2. Remove samples which have broad spectrum in RLE value plot
3. Remove samples which have strong deviation in RLE-NUSE mutiplot
4. Remove samples which exceed 99% line in RLE-NUSE T2 plot

By the exclusion criteria above, I removed GSM721017 from unstable angina samples and GSM217774 from healthy normal controls. Then, Genespring XI software was done to analysis the significant expressed genes between ARDS and healthy control leukocytes. P value cut-off point is less than 0.05. Fold change cut-off point is >2.0 fold change. Benjamini-hochberg corrected false discovery rate was used during the analysis. Totally, a genelist of 6020 genes was generated from the HGU133A2.0 chip with 18400 transcripts including 14500 well-characterized human genes.

## Results

### Toll-like receptor up-regulation in unstable angina

During the microarray analysis of peripheral blood leukocytes of unstable angina patients, I find out that Toll-like receptor genes are up-regulated.(Table1) These genes include TLR 1,2,4,6,8, TIRAP, TTRAP, TRAF6, TANK, TICAM2, and TOLIP. It is worth noting that TLR1 is 7 fold up-regulated, TLR4 is 13 fold up-regulated, TLR8 is 6 fold up-regulated, and TANK is 6 fold up-regulated. Since TLR 1,2,4,6,8 are related to the activation of antibacterial TH1 or TH17(TH22) immunity, we can say that TH1(anti-intracellular bacteria) or TH17(TH22)(anti-extracellular bacteria) immunity is triggered during unstable angina.

### Heat shock protein up-regulation in unstable angina

In the peripheral leukocyte of unstable angina patients, mRNAs of many heat shock proteins are up-regulated. These genes include HSPA6, HSPA1A/A1B, HSPA13, HSPB11, DNAJA4, DNAJB9, DNAJB2, DNAJB4, DNAJC6, DNAJA2, DNAJC27, DNAJC3, and DNAJC25. (Table2) Among them, HSPA1A/HSPA1B is 5.6 fold up-regulated and DNAJC6 is 4.8 fold up-regulated. However, few certain heat shock protein genes are down-regulated including HSP90AA1, DNAJC1, DNAJC2, DNAJC3, and DNAJC7. In addition, these genes are only 2-3 fold down-regulated. HSP70 can bind to TLRs to initiate anti-bacterial immunity. Thus, anti-bacterial TH1 or TH17(TH22) is likely to be activated at unstable angina.

TH1 and TH17 related transcription factor up-regulation during unstable angina

Then, we will look at TH1 and TH22(TH17) related transcription factor regulation during unstable angina.(Table 3) TH1 related transcription factors, STAT1, ETS2, RARA, RXRA, IRF2, MNDA, MYADM, IDO1, and MMD, are up-regulated during unstable angina. Up-regulated TH17 related transcription factors include SMAD1, SMAD4, SMAD2, SMAD5, SMAD7, RARA, RXRA, STAT5B, SOCS3, IKZF3, FOSL2, and FOXO3. Besides, negative regulator of TH17 immunity, ETS1, is down-regulated in unstable angina. In addition, THαβ related transcription factors are down-regulated including MAFB, suppressor of IKKe, and BCL6. STAT1 is the key downstream mediator of IFNg. ETS2 can up-regulate IL-12 production. Retinoic acid (for RXRA and RARA) can promote TH1 and TH17 immunity. IRF2 can promote TH1 immunity via IFNg. IDO is the downstream mediator of IFNg. MNDA, MYADM, and MMD are all macrophage differentiation factors. SMADs and STAT5B are downstream mediators of TGFb, a key cytokine of TH17 immunity. SOCS3, IKZF3, FOSL2, and FOXO3 can all promote TH17 immunity.

TH1 and TH17 related chemokine up-regulation during unstable angina

In this microarray analysis, we can see many chemokine genes are up-regulated during unstable angina. (Table 4) Most important of all, majority of these chemokine genes are TH1 and TH17 related chemokines. The up-regulated chemokine as well as chemokine receptors include CXCL1, CCR1, CXCL6, CCR2, CCR6, CCR3, DARC, CCL23, CCRL2, CXCR7, CXCL5, CKLF, IL8RB, IL8RA, IL8, S100A11, S100B, S100A9, S100P, and S100A12. Among these genes, CXCL1 is 8 fold up-regulated, CCR3 is 18 fold up-regulated, CXCL5 is 8 fold up-regulated, IL8RB is 38 fold up-regulated, and S100P is 50 fold up-regulated. TH1 related chemokine or chemokine receptors include CCR1, CCR2, CCR6, CCR3, CCL23, and CXCR7. TH17 related chemokine or chemokine receptors include CXCL1, CXCL6, CXCL5, IL8RB, IL8RA, IL8, and S100 proteins.

TH1 and TH17 related prostaglandin and leukotriene during unstable angina

In peripheral leukocytes of unstable angina patients, many prostaglandin and leukotriene genes are up-regulated. (Table 5) These up-regulated genes include HPGD, PTGS1, PTGS2, LTB4R, ALOX5AP, ALOX5, ALOX12, PLA2G12A, and PLAA. Prostaglandin and leukotriene B4 are mainly neutrophil chemoattractants. Thus, up-regulation of these genes suggests that TH17 related chemotaxis is up-regulated during unstable angina.

TH1 and TH17 related cytokine up-regulation in unstable angina

During unstable angina, many TH1 and TH17 related cytokines are up-regulated. (Table 6) Up-regulated TH1 cytokines include TGFA, IL-18(IFNG inducing factor), and IFNG. Up-regulated TH17 related cytokines include IL-8 and IL-1B. It is worth noting that IL1B is 11 fold up-regulated. Down-regulated cytokines in unstable angina include IL-34, IL-23A, IL-32, and TGFB1. It is important to know that IL1 beta is the key cytokine in TH17 (TH22) anti-extracellular bacteria immunity. IFN gamma is the key enzyme in TH1 anti-intracellular bacteria immunity. Thus, both TH17 and TH1 immunity are up-regulated during unstable angina.

TH1 and TH17 related cytokine receptor expression in unstable angina

In this microarray study, I find out that many cytokine receptor gene expressions are changed during unstable angina. (Table 7) Up-regulated cytokine is usually accompanying with down-regulated its cytokine receptor in certain specific immunological pathway. Down-regulation of IL-6 receptor (IL-6ST) implies that TH17 immunological pathway is activated. Down-regulation of IFNg receptor implies that TH1 immunological pathway is also activated. Up-regulation of IL-13 and IL-5 receptors suggest that TH2 immunological pathway is not activated. Up-regulation of IL-10 and type 1 interferon receptors suggests that THαβ immunological pathway is not activated. Significant up-regulation of IL-1 receptors suggests that TH22 immunological pathway is not activated. And, up-regulation of TGFB1 receptor suggests that Treg pathway is not activated.

Anti-bacterial complement up-regulation during unstable angina

Complements are important for anti-bacterial innate immunity including TH1 and TH17 immunity. Here, we show that majority of complement genes are up-regulated during unstable angina. (Table 8) These complement genes include CD59, CD55, CFD, C4BPA, CR1, C4A/B, CD46, C1QBP, ITGAX, C1RL, C5AR1, and CR1/1L. The only down-regulated complement gene is C1QA. It means that the whole complement machinery is up-regulated during unstable angina.

Other anti-bacterial related gene up-regulation in unstable angina

In table 9, many other important anti-bacterial genes are up-regulated during unstable angina. These genes include CSF3R, FPR1, CSF2RB, SCARF1, CSF2RA, DEFA4, NCF4, NCF2, FPR2, MRC2, DEFB106A, and PTX3. It is worth noting that CSF2RB is 10 fold up-regulated, DEFA4 is 11 fold up-regulated, and FPR2 is 38 fold up-regulated. These neutrophil or anti-bacterial related genes suggest the activation of TH1 and TH17 host immunological pathways in unstable angina.

Glycolytic genes up-regulation during unstable angina

In table 10, the whole set of glycolytic pathway is up-regulated in unstable angina. Hypoxia can drive the activation of the anaerobic glycolysis. Thus, it is reasonable that glycolytic enzymes are up-regulated during the hypoxia status of the attack of unstable angina. These up-regulated genes include PGK1, PFKFB3, HK2, PYGL, BPGM, PDK3, PFKFB2, GAPDH, ENO1, PDK2, PDK4, PGK1, and PFKFB4. It is worth noting that BPGM(2,3-bisphosphoglycerate mutase) is 19 fold up-regulated. The product of BPGM is 2,3-bisphosphoglycerate, which combines with hemoglobin, can cause a decrease in affinity for oxygen. Thus, the presence of 2,3-bisphosphoglycerate helps oxyhemoglobin to unload oxygen to help unstable angina patients.

$H^+$-ATPase gene up-regulation during unstable angina

In table 11, we can see many $H^+$-ATPases are up-regulated during unstable angina. These genes include ATP6V0B, ATP6V0C, ATP6V1B2, ATP6V0E1, ATP6AP2, ATP6V1A, ATP6V1C1, ATP6V0A2, and ATP6V1D. These ATPases are proton pumps which can generate $H^+$ by using ATP. In addition, several carbonic anhydrases are also up-regulated including CA1, CA2, and CA4. Carbonic anhydrases can catalyze $H_2CO_3$ formation from $CO_2$ and $H_2O$. Thus, acidosis can result during the attack of unstable angina.

Coagulation gene up-regulation during unstable angina

In table 12, majority of coagulation related genes are up-regulated in peripheral leukocytes of unstable angina patients. These genes include THBS1, VWF, THBD, F2R,

F5, F8, F2RL1, ITGA2B, PROS1, GP5, PTAFR, PLAUR, TFP1, ITGB3, PDGFD, HPSE, GP6, PEAR1, and TBXAS1. Only few coagulation related genes are down-regulated including GP1BB and CD36. Most important of all, F2RL1 is 18 fold up-regulated.

RBC related genes are up-regulated during unstable angina

In table 13, majority of RBC related genes are up-regulated during unstable angina. These up-regulated genes include EPB41, GYPE, ANK1, HP/HPP, NFE2, ALAS2, GYPA, HBG1/HBG2, EPOR, RHCE/RHD, ANKRD12, GYPB, HEBP1, ERAF, HBQ1, AK2RIN2, ANKRDS2, HEMGN, ANKRD50, HBM, and HBD. A few genes are down-regulated including HMOX1 and ANKRD10. It is worth noting that many hemoglobin genes are up-regulated including HBG1, HBG2, HBQ1, HBM, and HBD. Among them, HBG1/2 is 20 fold up-regulated, HBM is 60 fold up-regulated, HEMGN is 49 fold up-regulated, HBQ1 is 18 fold up-regulated, ERAF is 33 fold up-regulated, GYPA is 27 fold up-regulated, and ALAS2 is 36 fold up-regulated. Heme oxygenase (HMOX1) can degrade heme and it is down-regulated in unstable angina.

Discussion

Atherosclerosis is a very common and detrimental disease, especially in many developed countries. It is usually among the top three leading causes of mortality. Inflammatory process is thought to be related to the etiology of atherosclerosis. C reactive protein (CRP) is found to be related to the course of atherosclerosis. Abnormal T cell activation has been found in atherosclerosis.(Liuzzo, Kopecky et al. 1999) However, detailed relationship between inflammation and atherosclerosis is remained to be unlocked. Unstable angina is a clinical manifested syndrome of atherosclerosis. Thus, I use microarray analysis of peripheral leukocytes to study the relationship of specific inflammation and unstable angina.

Host immunity against bacteria includes TH1(anti-intracellular bacteria) and TH17(anti-extracellular bacteria). It is observed that Chlamydia pneumonia is noted in endothelium of atherosclerosis lesion(Benagiano, Azzurri et al. 2003, Benagiano, Munari et al. 2012). Thus, immunity against intracellular bacteria is likely to be activated in atherosclerosis. In addition, Helicobacter pylori infection is also related to atherosclerosis. Thus, anti-extracellular TH17 immunity is also likely to be activated during atherosclerosis(Ng, Burris et al. 2011). In addition, TH1 and TH17 immunological pathways are also related to important risk factors for atherosclerosis: Diabetes mellitus and hypertension.(Gao, Jiang et al. 2010, Xie, Wang et al. 2010) In addition, CRP, the TH17 inflammatory factor, is elevated in atherosclerosis. Macrophage, the main TH1 effector cell, is accumulated in atherosclerotic plaques(Laurat, Poirier et al. 2001). In addition, deficiency of Treg cells promotes the progression of atherosclerosis.(Gotsman, Grabie et al. 2006, Mor, Planer et al. 2007) Besides, TH2 and THαβ immunological pathway activation such as IL-10 up-regulation can reduce the development of atherosclerosis.(Liuzzo, Kopecky et al. 1999, Pinderski 2002) These evidences all suggest that TH1 and TH17 immunity plays an important role in the pathophysiology of atherosclerosis.

In this study, I find out the evidences that immunity plays a very important role in the pathogenesis of unstable angina. First of all, anti-bacterial Toll-like receptors are up-regulated during unstable angina. These Toll-like receptors include TLR 1,2,4,6,8. It is worth noting that TLR 1,2,4,6 are for triggering anti-extracellular bacteria TH17 immunity. TLR8 is for triggering anti-intracellular bacteria TH1 immunity. Thus, both TH1 and TH17 immunity are initiated during unstable angina.

In addition, heat shock proteins which are responsible for triggering anti-bacterial immunity are also up-regulated during unstable angina attack. Heat shock protein 70 (HSP70) can bind to Toll-like receptor 2 and Toll-like receptor 4 to trigger TH17 immunity. In this study, I find out that HSP70 genes are up-regulated including HSPA1A, HSPA6, and HSPA13. This means anti-bacterial immunity is activated during unstable angina.

Then, we look at immune-related transcription factor expression in unstable angina. Strikingly, we find out many TH1 and TH17 related immune transcription factor up-regulation in unstable angina patients' peripheral leukocytes. STAT1 is the key downstream transcription factor of TH1 immunity. The key TH1 cytokine, IFN gamma, can mediate its function mainly by activating STAT1. In addition, STAT5B and SMAD proteins are major effector genes of TGF beta signaling. TGF beta is the key cytokine to initiate TH17 immunity. Thus, both TH1 and TH17 are likely to be activated during unstable angina. In addition, TH17 negative regulator, ETS1, is down-regulated in unstable angina(Moisan, Grenningloh et al. 2007). Retinoic acid related transcription factors, which promote TH1 and TH17 immunity, are up-regulated during unstable angina.(Mucida, Park et al. 2007) Aiolos(IKZF3), which can activate TH17 immunity, is also up-regulated in unstable angina.(Quintana, Jin et al. 2012)

Chemokine genes are differentially regulated in unstable angina. Most important of all, TH17 and TH1 related chemokine and chemokine receptor genes are up-regulated. These strikingly up-regulated chemokine related genes include CXCL1, CCR3, CXCL5, IL8RB, and S100P. CCR3 is TH1 immunity related chemokine receptor. CXCL1, CXCL5, and S100P are TH17 immunity related chemokines. And, IL8RB is TH17 related chemokine receptor. Besides, prostaglandin and leukotriene B4 related genes are also up-regulated during unstable angina. This also shows that TH17 anti-bacterial immunity is activated at unstable angina.

Then, we look at cytokine and cytokine receptor gene expression profiles of peripheral leukocytes during unstable angina. We find out that IL-18, TGFA, IFNG, IL1B, and IL8 are up-regulated. Among them, IL-18, TGFA, and IFNG are key TH1 related cytokines. IL8 and IL1B are key TH17 related cytokines. Besides, many cytokine receptors are also differentially regulated at unstable angina. These up-regulated cytokine receptors include IL-1R, IL-13R, IL-5R, IL10R, IFNaR, IL1R, and TGFB1R. And, down-regulated cytokine receptors are IL6ST and IFNgR. Specific immunological pathway is usually associated with down-regulation of its effector cytokine receptors. Thus, TH1 activation is related to interferon gamma receptor down-regulation, and TH17 activation is related to interleukin 6 receptor down-regulation. In addition, TH2, THαβ, Treg, and TH22 immune pathways are not triggered. These results also support my hypothesis.

Complements are important in mediating host killing of intracellular bacteria as well as extracellular bacteria. In this study, we can see that majority of complement machinery is activated including CD59, CD55, CFD, C4BPA, CR1, C4A/B, CD46, C1QBP, ITGAX, C1RL, C5AR1, and CR1/1L. This means anti-bacterial TH1 or TH17 immunity is triggered during unstable angina. Besides, other important anti-bacterial host defense genes are also up-regulated. These up-regulated anti-bacterial genes include CSF3R, FPR1, CSF2RB, SCARF1, CSF2RA, DEFA4, NCF4, NCF2, FPR2, MRC2, DEFB106A, and PTX3. These genes include neutrophil and macrophage growth factors, pattern recognition receptor, defensin, neutrophil cytosolic factors, and pentraxin. These all suggest that TH1 or TH17 anti-bacterial immunity is activated during unstable angina.

By using microarray analysis, we can also understand the complications of unstable angina. Glycolysis, acidosis, hypoxia, and coagulation anomaly are frequently reported in unstable angina. First of all, we look at the glycolytic mediating enzymes in unstable angina. Strikingly, we find out all glycolytic enzyme genes are up-regulated including PGK1, PFKFB3, HK2, PYGL, BPGM, PDK3, PFKFB2, GAPDH, ENO1, PDK2, PDK4, PGK1, and PFKFB4. Among them, BPGM is 19 fold up-regulated. BPGM can help to unload oxygen from oxy-hemoglobins. Thus, it can help to alleviate the hypoxic syndrome during unstable angina. It can be explained as a host compensatory mechanism responding to the attack of unstable angina. More, we find out that many ATPase genes are up-regulated during unstable angina including ATP6V0B, ATP6V0C, ATP6V1B2, ATP6V0E1, ATP6AP2, ATP6V1A, ATP6V1C1, ATP6V0A2, and ATP6V1D. Several carbonic anhydrase enzyme genes are also up-regulated including CA1, CA2, and CA4. In my previous malaria genomic research, I find out a co-expression of glycolytic enzymes and ATPases. Here, I also find out acidosis at unstable angina can be due to up-regulation of these proton pumps ($H^+$-ATPase) and $H_2CO_3$ producing carbonic anhydrases.

Finally, we look at RBC and platelet related gene expression profiles during unstable angina. Strikingly, we find out that majority of RBC and platelet related genes are up-regulated during unstable angina. Up-regulated platelet/coagulation related genes include THBS1, VWF, THBD, F2R, F5, F8, F2RL1, ITGA2B, PROS1, GP5, PTAFR, PLAUR, TFP1, ITGB3, PDGFD, HPSE, GP6, PEAR1, and TBXAS1. Platelet function is triggering blood coagulation, and overactivation of coagulation related genes can be related to the pathophysiology of atherosclerosis induced unstable angina with coronary artery occlusion. Up-regulated RBC related genes include EPB41, GYPE, ANK1, HP/HPP, NFE2, ALAS2, GYPA, HBG1/HBG2, EPOR, RHCE/RHD, ANKRD12, GYPB, HEBP1, ERAF, HBQ1, AK2RIN2, ANKRDS2, HEMGN, ANKRD50, HBM, and HBD. We can see many hemoglobin genes are up-regulated at unstable angina. Unstable angina is caused by lack of oxygen support for cardiac muscle. Thus, up-regulated RBC related genes including hemoglobin genes can help to alleviate the hypoxia status during unstable angina. It can be viewed as a host compensatory mechanism at the attack of unstable angina.

In summary, microarray analysis of peripheral leukocytes from unstable angina patients can help to explain the pathogenesis of unstable angina. From the above evidences, unstable angina can be viewed as a mix TH1 and TH17 inflammatory disorder with up-regulated TH1/TH17 chemokines, cytokines, transcription factors, Toll-like receptors, heat shock proteins, complements, leukotrienes, prostaglandins, defensins, pentraxins, and other effector molecules. In addition, overactivation of platelet related genes, glycolytic enzymes, and $H^+$-ATPases can help to explain the pathophysiology of unstable angina with coagulation hyperactivity and metabolic acidosis. Finally, up-regulation of RBC related genes should be a host compensatory mechanism to increase oxygen delivery during unstable angina.

**Author's information**

Wan-Chung Hu is a MD from College of Medicine of National Taiwan University and a PhD from vaccine science track of Department of International Health of Johns Hopkins University School of Public Health. He is a postdoctorate in Genomics Research Center of Academia Sinica, Taiwan. His previous work on immunology and functional genomic studies were published at *Infection and Immunity* 2006, 74(10):5561, *Viral Immunology* 2012, 25(4):277, and *Malaria Journal* 2013,12:392. He proposed THαβ immune response as the host immune response against viruses.

Figure legends

Figure 1. RMA express plot for selecting samples in normal healthy controls.
1-A NUSE boxplot for normal control
1-B RLE boxplot for normal control
1-C RLE-NUSE multiplot for normal control
1-D RLE-NUSE T2 plot for normal control

Figure 2. RMA express plot for selecting samples in unstable angina patients.
2-A NUSE boxplot for unstable angina patients
2-B RLE boxplot for unstable angina patients
2-C RLE-NUSE multiplot for unstable angina patients
2-D RLE-NUSE T2 plot for unstable angina patients

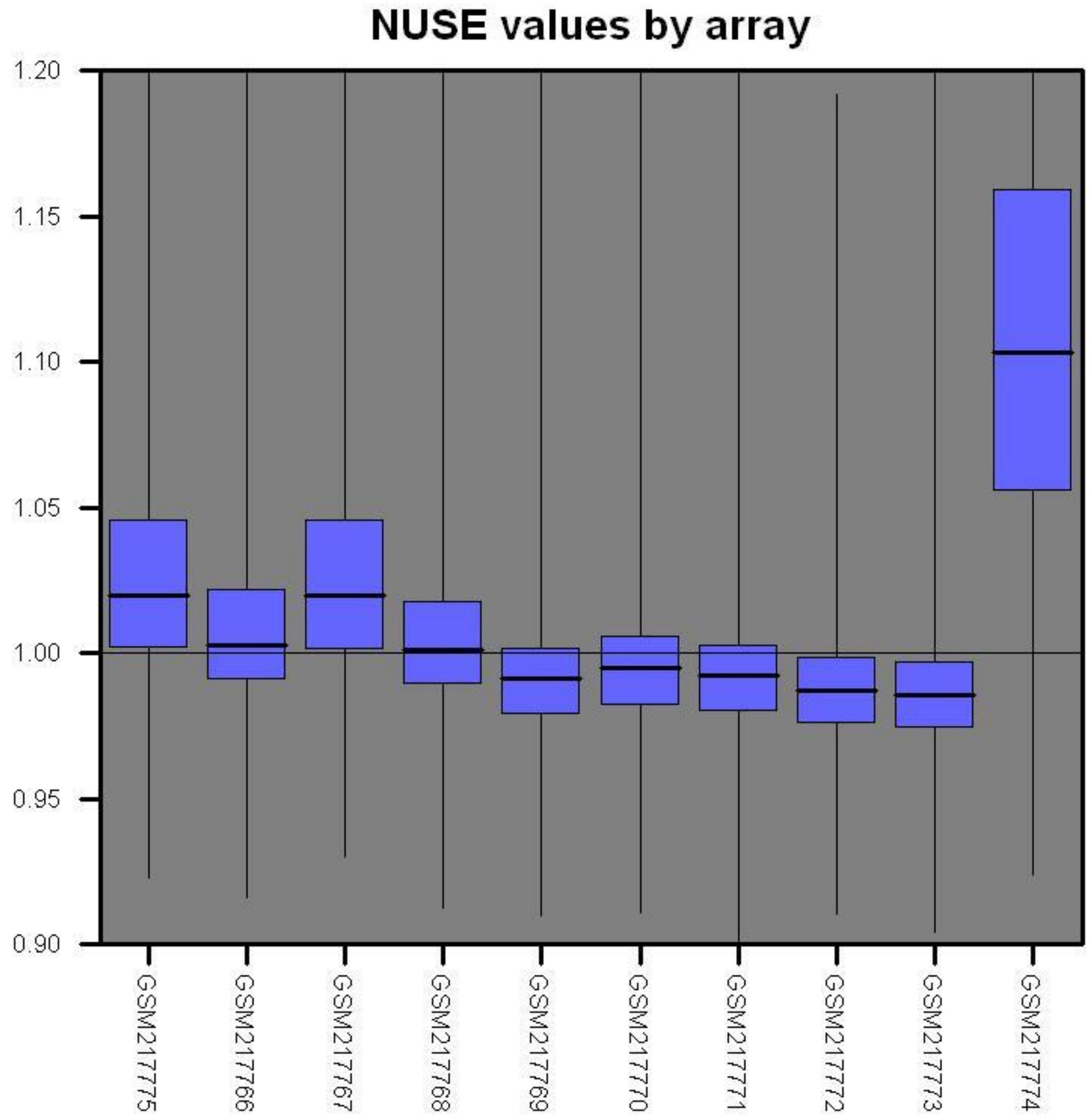

Figure 1-A

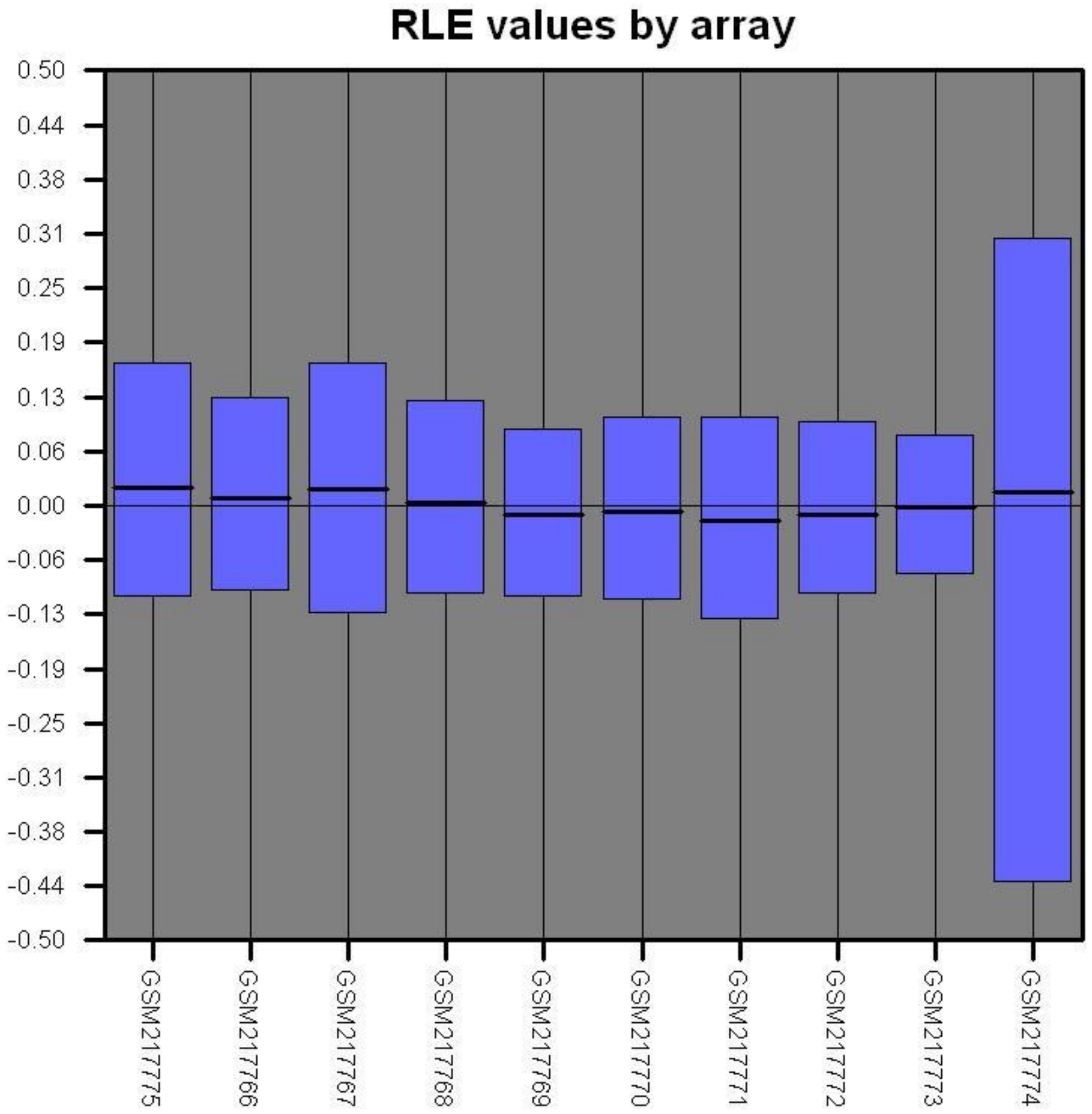

Figure 1-B

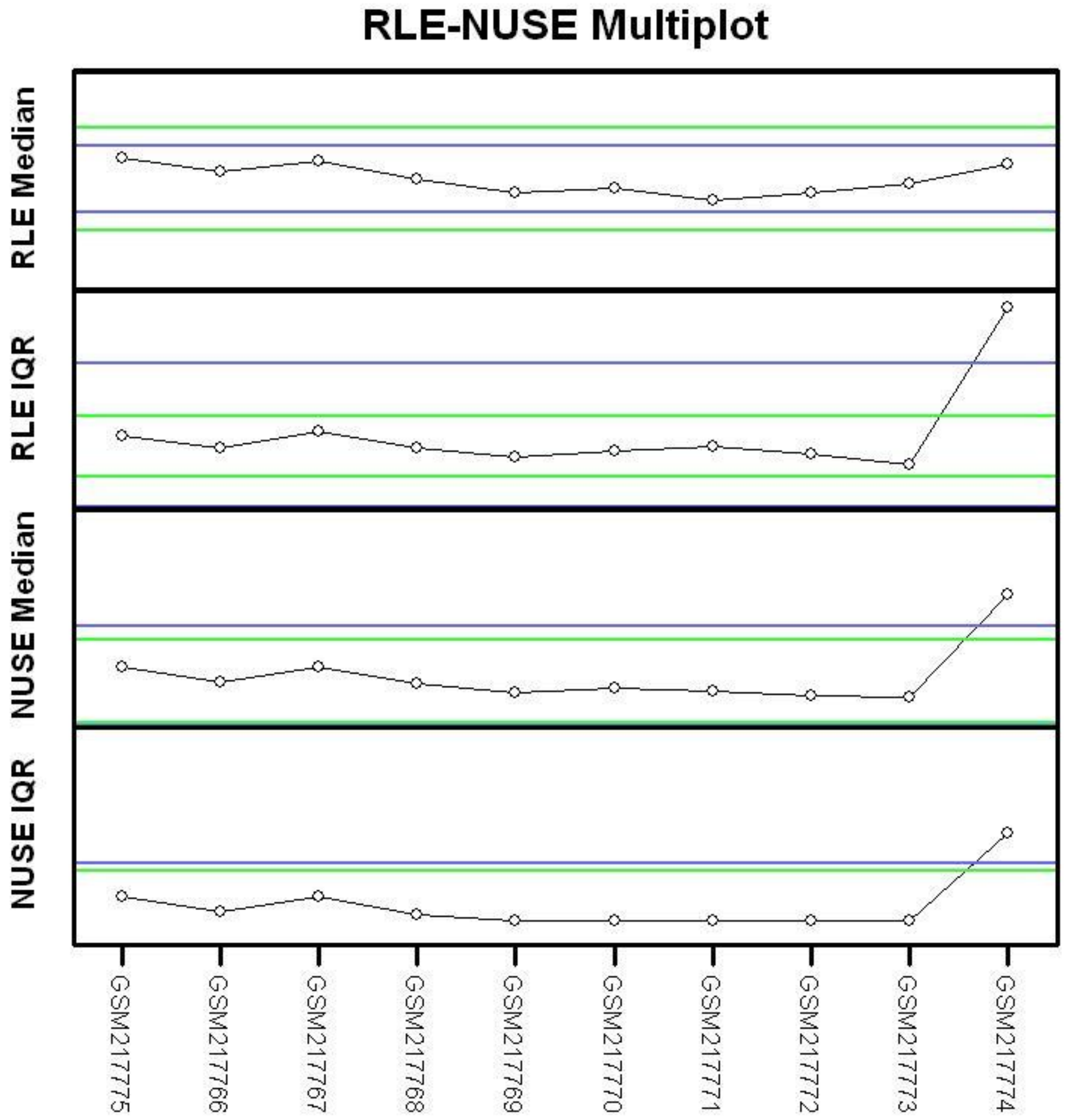

Figure 1-C

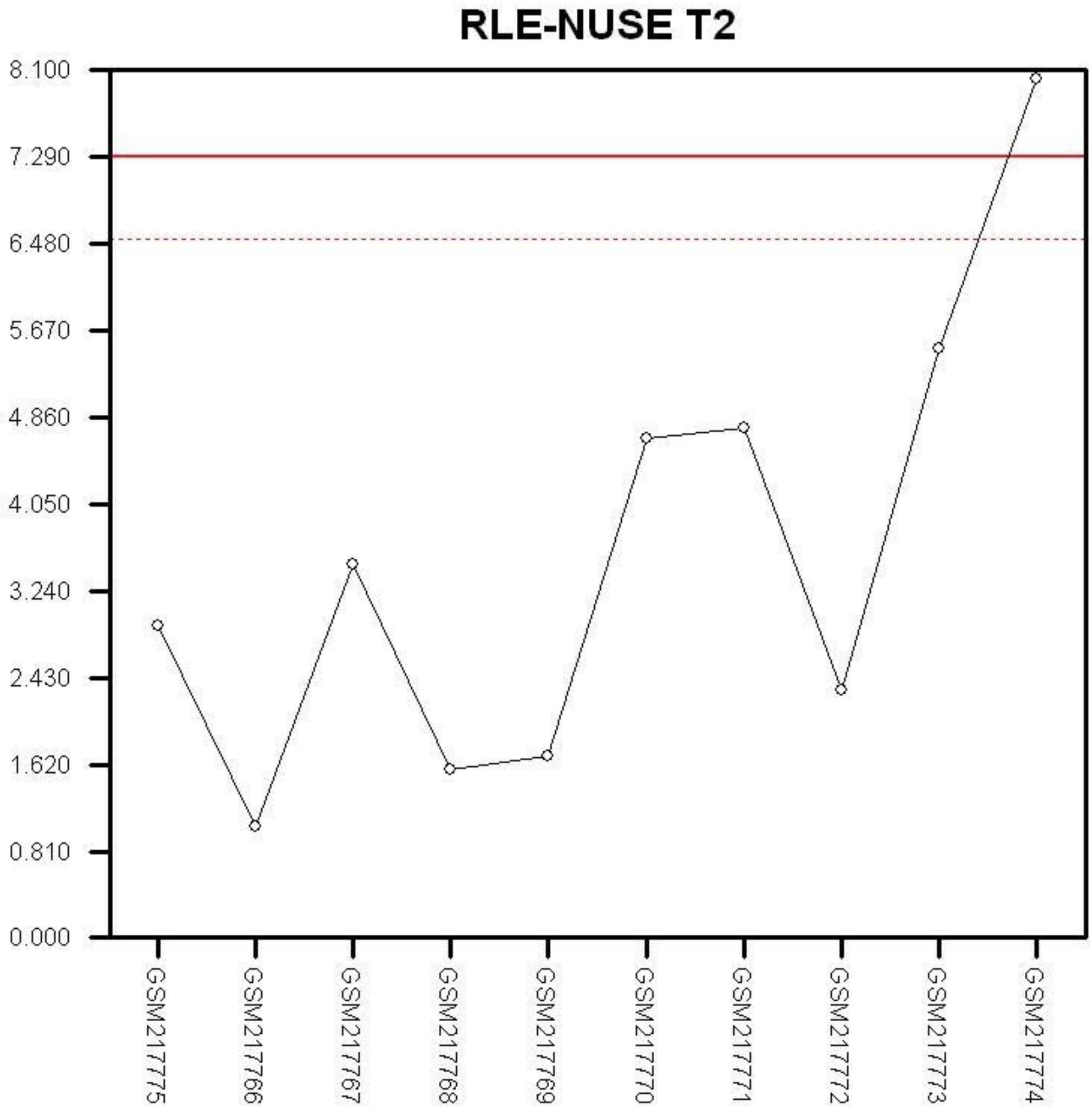

Figure 1-D

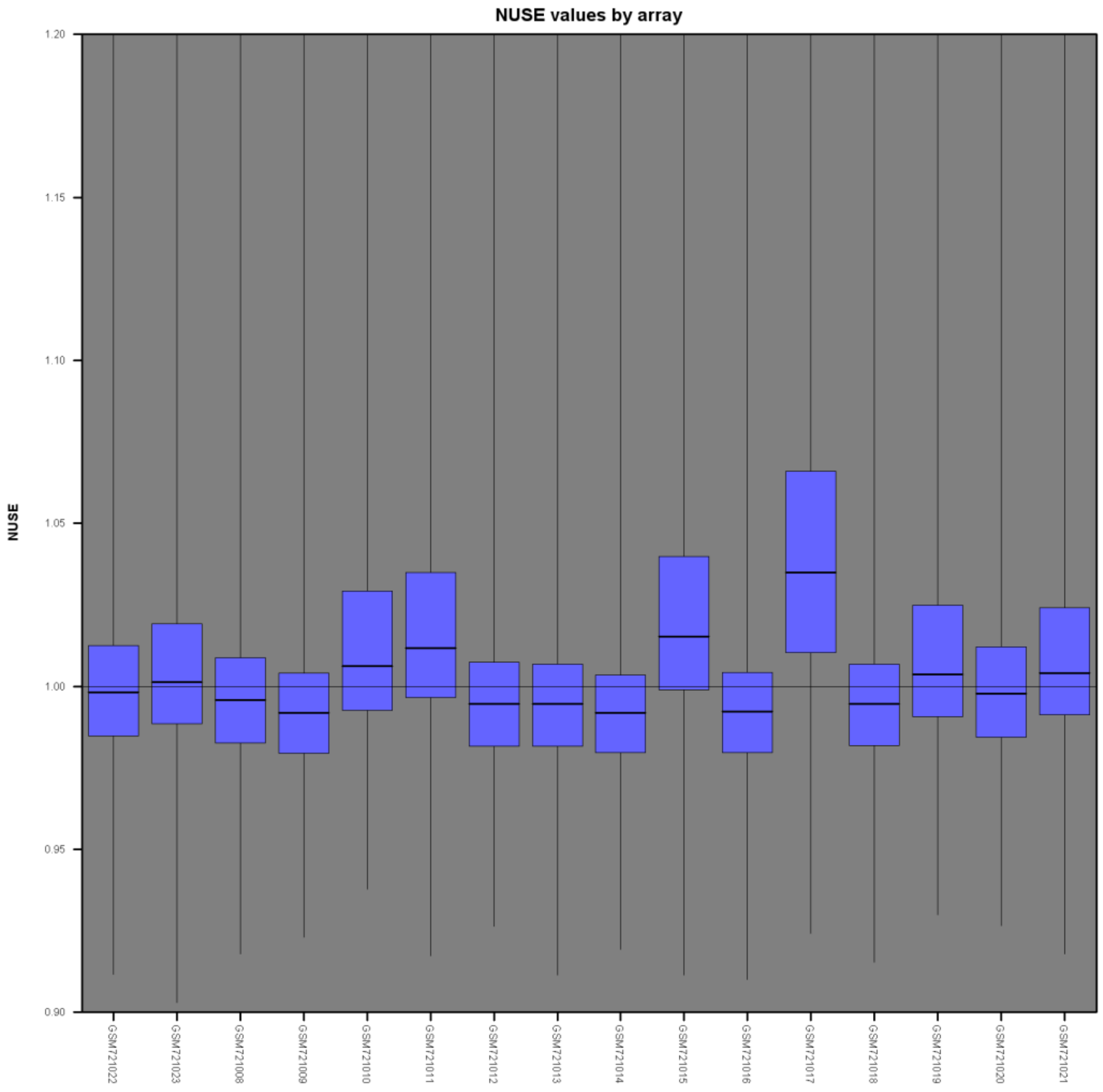

Figure 2-A

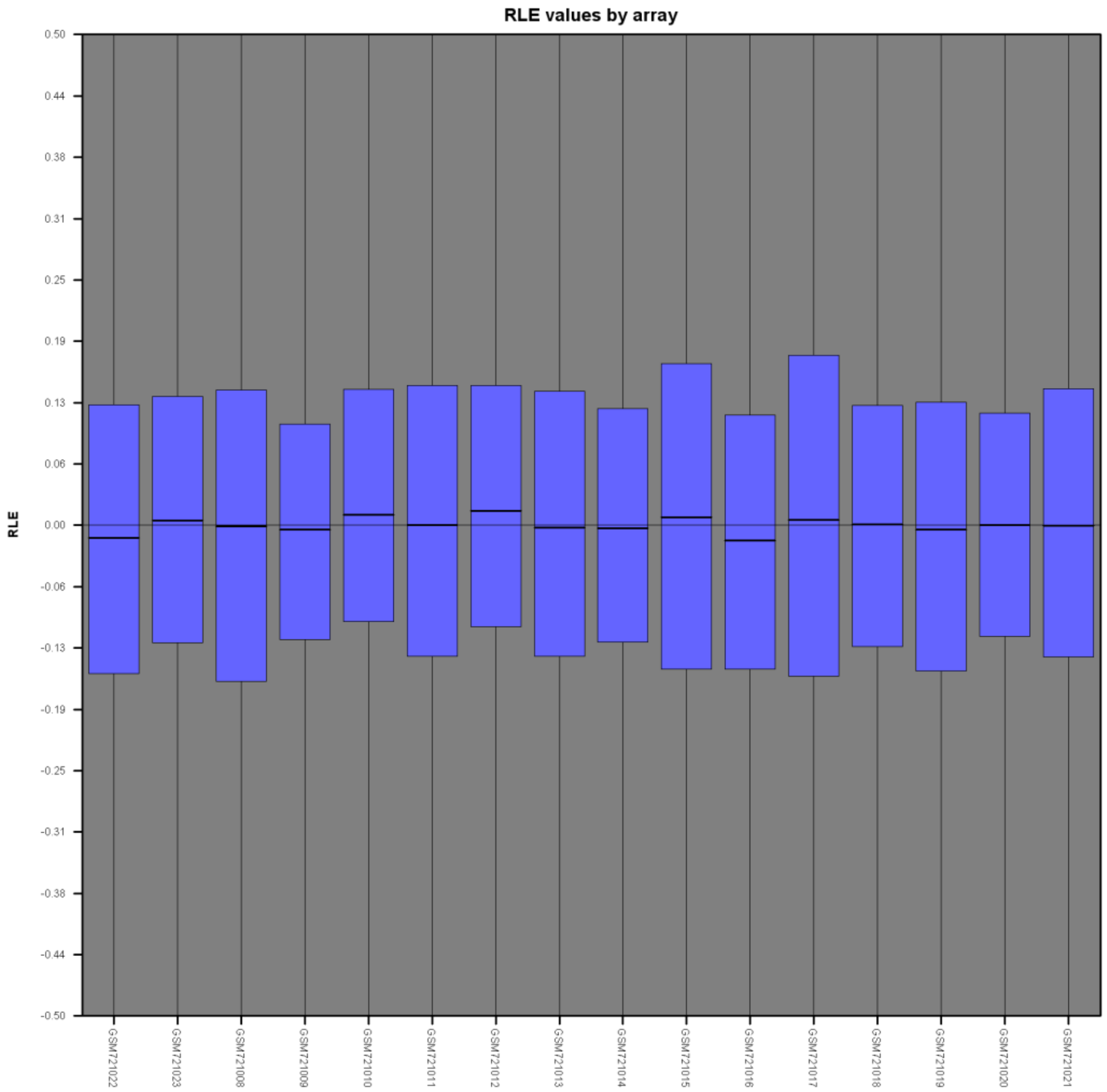

Figure 2-B

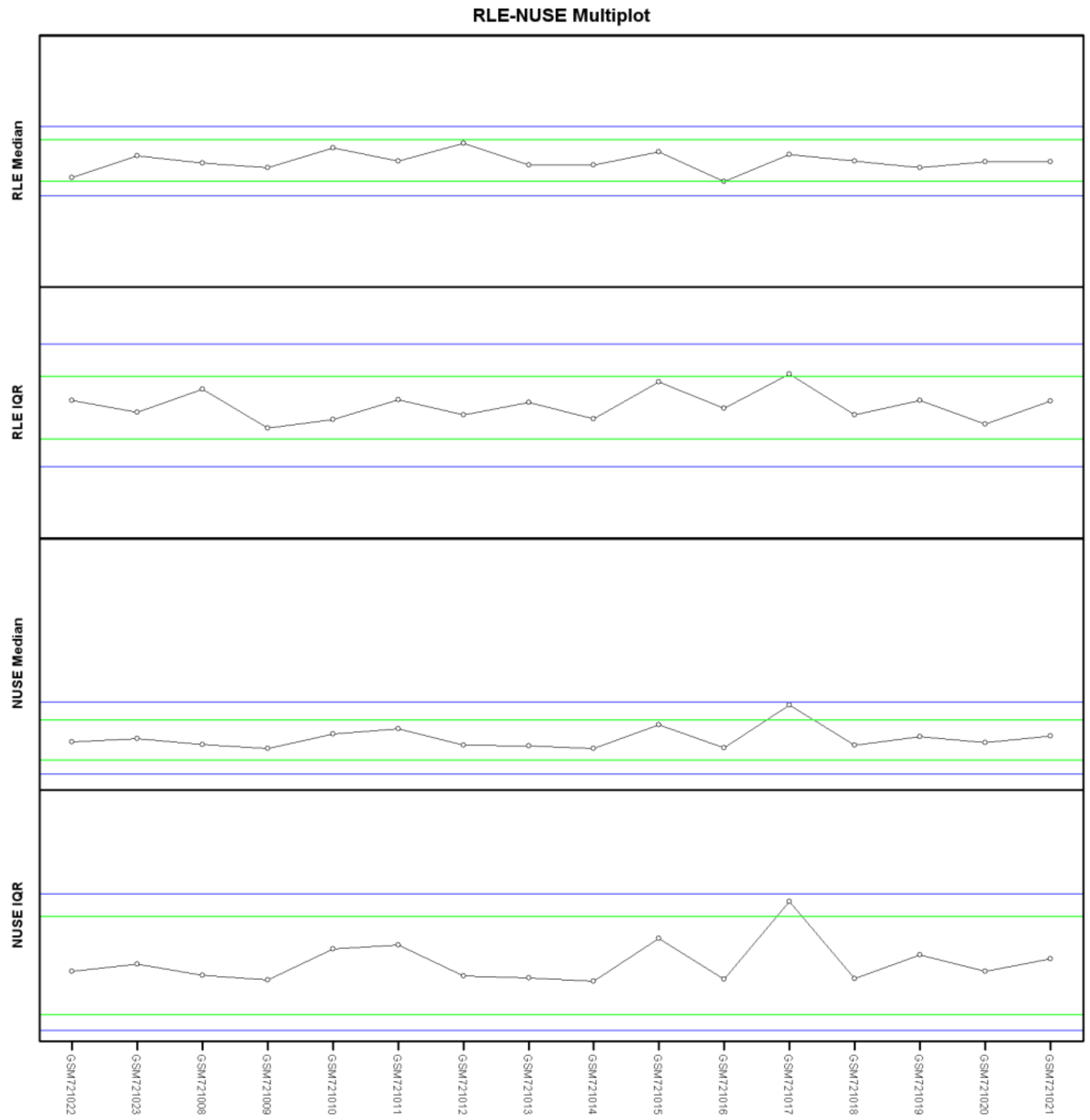

Figure 2-C

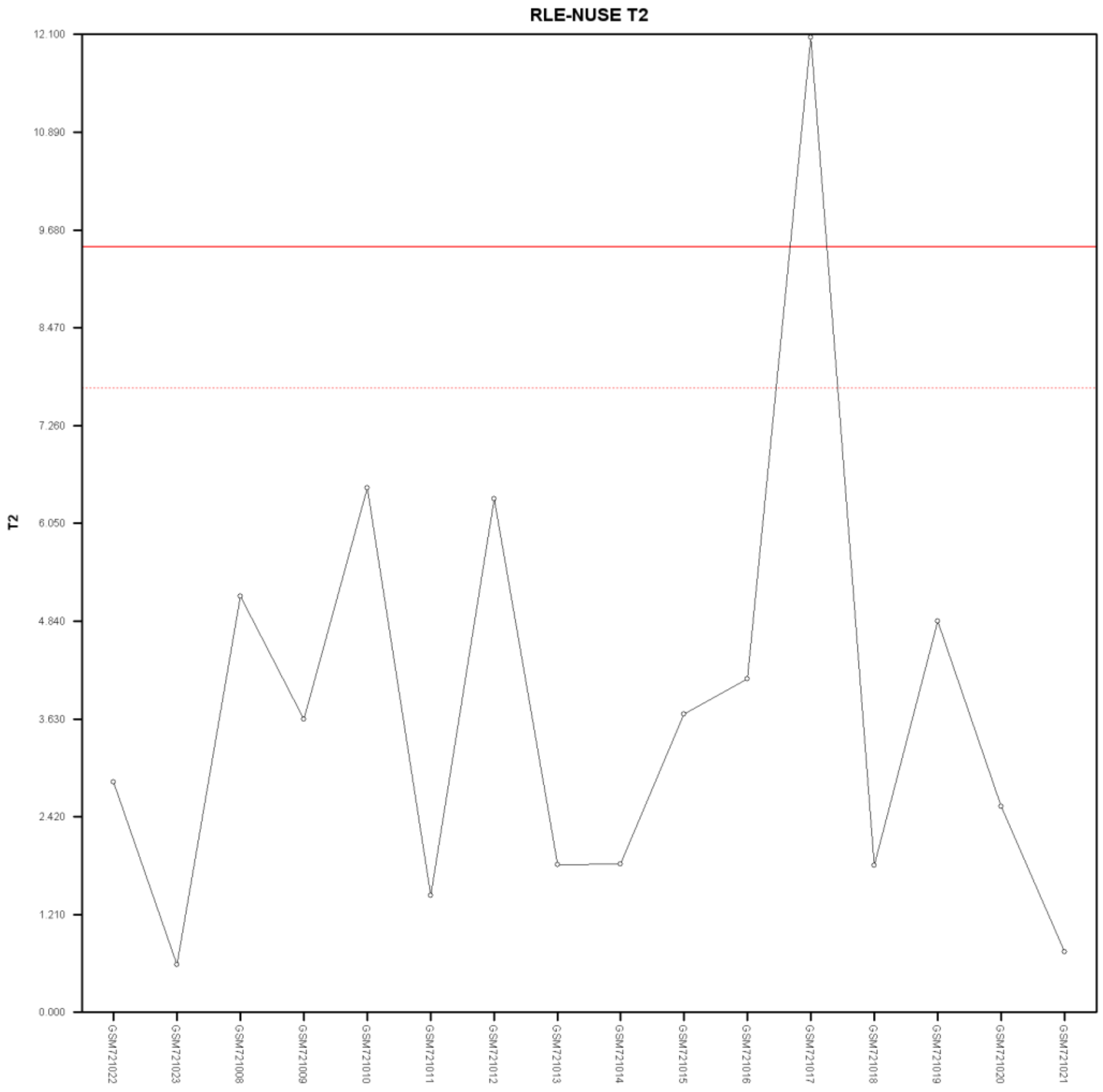

Figure 2-D

Table 1.

Toll-like-receptor

| Probe Set ID | Pvalue | Arrow | Fold | Gene |
|---|---|---|---|---|
| 204924_at | 6.07E-05 | up | 2.711394 | TLR2 |
| 207446_at | 1.82E-06 | up | 2.196196 | TLR6 |
| 210176_at | 1.61E-09 | up | 7.237305 | TLR1 |
| 220832_at | 1.16E-08 | up | 6.489917 | TLR8 |
| 221060_s_at | 6.35E-07 | up | 5.706266 | TLR4 |
| 224341_x_at | 5.14E-06 | up | 5.151984 | TLR4 |
| 229560_at | 6.43E-05 | up | 3.756611 | TLR8 |
| 232068_s_at | 2.04E-10 | up | 13.17082 | TLR4 |
| 1552798_a_at | 7.63E-07 | up | 3.467543 | TLR4 |
| 202266_at | 1.91E-07 | up | 2.636574 | TTRAP |
| 209451_at | 5.11E-09 | up | 5.10644 | TANK |
| 210458_s_at | 2.56E-10 | up | 6.495119 | TANK |
| 211828_s_at | 1.66E-08 | down | -2.46097 | TNIK |
| 217930_s_at | 8.09E-10 | up | 3.414113 | TOLLIP |
| 239431_at | 1.01E-05 | up | 2.18973 | TICAM2 |
| 1552360_a_at | 1.28E-09 | up | 2.909597 | TIRAP |
| 213817_at | 6.28E-04 | up | 2.556772 | IRAK3 |
| 205558_at | 2.29E-09 | up | 2.273412 | TRAF6 |

Table 2. Heat shock protein

| Probe Set ID | Pvalue | Arrow | Fold | Gene |
|---|---|---|---|---|
| 211968_s_at | 5.02E-09 | down | -3.03611 | HSP90AA1 |
| 213418_at | 1.09E-04 | up | 2.462545 | HSPA6 |
| 117_at | 2.01E-06 | up | 3.369852 | HSPA6 |
| 200799_at | 1.78E-05 | up | 2.131275 | HSPA1A |
| 200800_s_at | 7.79E-05 | up | 2.27759 | HSPA1A/B |
| 202557_at | 1.48E-05 | up | 2.020725 | HSPA13 |
| 202581_at | 7.23E-10 | up | 5.616055 | HSPA1A/B |
| 203960_s_at | 1.04E-05 | up | 2.114942 | HSPB11 |
| 1554333_at | 4.79E-05 | up | 4.757214 | DNAJA4 |
| 1554334_a_at | 2.40E-04 | up | 3.128753 | DNAJA4 |
| 1554462_a_at | 6.27E-08 | up | 2.534985 | DNAJB9 |
| 1558080_s_at | 2.64E-06 | down | -3.33239 | DNAJC3 |
| 202416_at | 1.77E-09 | down | -2.93756 | DNAJC7 |
| 202500_at | 6.74E-09 | up | 3.785045 | DNAJB2 |
| 202842_s_at | 1.22E-06 | up | 2.269558 | DNAJB9 |
| 202843_at | 5.11E-09 | up | 2.767084 | DNAJB9 |
| 203810_at | 4.20E-08 | up | 2.99098 | DNAJB4 |
| 204720_s_at | 2.13E-06 | up | 4.809938 | DNAJC6 |
| 208810_at | 8.05E-09 | up | 2.512607 | DNAJB6 |
| 208811_s_at | 1.66E-06 | up | 2.526036 | DNAJB6 |
| 209015_s_at | 2.35E-07 | up | 2.843226 | DNAJB6 |
| 209157_at | 3.03E-08 | up | 2.647895 | DNAJA2 |
| 213097_s_at | 3.59E-10 | down | -2.68168 | DNAJC2 |
| 222620_s_at | 5.74E-07 | down | -2.89806 | DNAJC1 |
| 223504_at | 2.84E-05 | up | 2.296595 | DNAJC27 |
| 225061_at | 1.83E-04 | up | 3.17514 | DNAJA4 |
| 225284_at | 4.93E-05 | up | 2.162836 | DNAJC3 |
| 226859_at | 5.59E-09 | up | 2.031987 | DNAJC25 |
| 226994_at | 2.89E-07 | up | 2.197155 | DNAJA2 |

Table 3. Transcription factors

| Probe Set ID | Pvalue | Arrow | Fold | Gene |
|---|---|---|---|---|
| M97935_MB_at | 5.29E-06 | up | 3.250345 | STAT1 |
| 201328_at | 8.13E-06 | up | 2.391284 | ETS2 |
| 201329_s_at | 2.07E-04 | up | 2.834679 | ETS2 |
| 202527_s_at | 2.47E-08 | up | 3.925728 | SMAD4 |
| 203077_s_at | 2.05E-09 | up | 2.499254 | SMAD2 |
| 203140_at | 2.67E-08 | up | 4.672891 | BCL6 |
| 203275_at | 3.53E-10 | up | 2.53243 | IRF2 |
| 203749_s_at | 1.45E-09 | up | 2.776926 | RARA |
| 204131_s_at | 3.18E-07 | up | 5.362668 | FOXO3 |
| 204132_s_at | 7.42E-10 | up | 9.790864 | FOXO3/B |
| 204790_at | 3.32E-06 | up | 2.693902 | SMAD7 |
| 205841_at | 8.44E-05 | up | 2.228473 | JAK2 |
| 210655_s_at | 2.51E-08 | up | 6.194851 | FOXO3/B |
| 211027_s_at | 1.09E-07 | down | -2.20832 | IKBKB |
| 212550_at | 7.95E-10 | up | 2.703492 | STAT5B |
| 214447_at | 2.18E-06 | down | -2.71117 | ETS1 |
| 217804_s_at | 7.97E-10 | down | -2.29951 | ILF3 |
| 218880_at | 2.81E-07 | up | 2.718907 | FOSL2 |
| 222670_s_at | 7.29E-04 | down | -2.03266 | MAFB |
| 223287_s_at | 8.80E-09 | down | -4.59015 | FOXP1 |
| 224889_at | 1.75E-10 | up | 8.903059 | FOXO3 |
| 224891_at | 3.03E-07 | up | 4.554951 | FOXO3 |
| 225223_at | 1.74E-06 | up | 2.836291 | SMAD5 |
| 225262_at | 0.018247 | up | 2.013291 | FOSL2 |
| 227798_at | 1.70E-05 | up | 2.374207 | SMAD1 |
| 228026_at | 1.18E-06 | up | 3.017289 | RP5-1000E10.4 |
| 228188_at | 5.50E-07 | up | 4.339457 | FOSL2 |
| 228758_at | 1.45E-05 | up | 3.365109 | BCL6 |
| 240613_at | 4.95E-10 | down | -6.16447 | JAK1 |
| 1552611_a_at | 2.95E-05 | up | 2.774677 | JAK1 |
| 1565703_at | 1.20E-06 | up | 2.179628 | SMAD4 |
| 204959_at | 1.12E-06 | up | 5.577443 | MNDA |
| 210029_at | 1.19E-05 | up | 3.981845 | IDO1 |
| 211027_s_at | 1.09E-07 | down | -2.20832 | IKBKB |
| 214105_at | 7.95E-08 | down | -2.10317 | SOCS3 |
| 221092_at | 8.50E-04 | up | 2.294967 | IKZF3 |
| 223287_s_at | 8.80E-09 | down | -4.59015 | FOXP1 |
| 223937_at | 5.84E-07 | down | -2.23877 | FOXP1 |
| 225223_at | 1.74E-06 | up | 2.836291 | SMAD5 |
| 225673_at | 1.62E-06 | up | 2.660096 | MYADM |
| 202426_s_at | 5.52E-11 | up | 7.436708 | RXRA |
| 228026_at | 1.18E-06 | up | 3.017289 | RP5-1000E10.4 |
| 235444_at | 2.78E-08 | down | -2.21648 | FOXP1 |
| 244523_at | 7.85E-07 | up | 2.832237 | MMD |

Table 4 Chemokine

| Probe Set ID | Pvalue | Arrow | Fold | Gene |
|---|---|---|---|---|
| 204470_at | 4.05E-11 | up | 7.821242 | CXCL1 |
| 205098_at | 1.50E-05 | up | 3.999256 | CCR1 |
| 205099_s_at | 0.001392 | up | 2.339561 | CCR1 |
| 206336_at | 4.69E-10 | up | 3.787778 | CXCL6 |
| 206337_at | 2.90E-06 | down | -2.92756 | CCR7 |
| 206978_at | 0.011513 | up | 2.081589 | CCR2 |
| 206983_at | 1.38E-04 | up | 2.189831 | CCR6 |
| 207794_at | 0.00177 | up | 2.296914 | CCR2 |
| 208304_at | 4.74E-11 | up | 17.6996 | CCR3 |
| 208335_s_at | 2.21E-06 | up | 3.161269 | DARC |
| 210548_at | 0.001164 | up | 2.080932 | CCL23 |
| 211434_s_at | 1.17E-05 | up | 2.202749 | CCRL2 |
| 212977_at | 8.10E-06 | up | 2.406206 | CXCR7 |
| 214974_x_at | 2.32E-06 | up | 7.833923 | CXCL5 |
| 215101_s_at | 2.78E-06 | up | 8.043318 | CXCL5 |
| 219161_s_at | 4.00E-11 | up | 5.634878 | CKLF |
| 221058_s_at | 1.44E-08 | up | 2.639918 | CKLF |
| 223451_s_at | 7.77E-11 | up | 4.474447 | CKLF |
| 223454_at | 4.20E-06 | up | 3.074875 | CXCL16 |
| 1568934_at | 2.01E-06 | up | 2.647086 | CX3CR1 |
| 204470_at | 4.05E-11 | up | 7.821242 | CXCL1 |
| 207008_at | 1.99E-12 | up | 38.26742 | IL8RB |
| 207094_at | 1.25E-14 | up | 12.5119 | IL8RA |
| 211506_s_at | 3.18E-04 | up | 2.965578 | IL8 |
| 202859_x_at | 0.02858 | up | 2.698248 | IL8 |
| 208540_x_at | 2.08E-06 | up | 2.709116 | S100A11/P |
| 209686_at | 0.022845 | up | 2.555998 | S100B |
| 238909_at | 5.34E-06 | down | -2.86673 | S100A10 |
| 203535_at | 4.97E-06 | up | 3.946168 | S100A9 |
| 204351_at | 5.64E-10 | up | 50.28864 | S100P |
| 205863_at | 1.63E-04 | up | 4.381631 | S100A12 |

Table 5 Prostaglandin/leukotriene

| Probe Set ID | Pvalue | Arrow | Fold | Gene |
|---|---|---|---|---|
| 211548_s_at | 4.79E-08 | up | 3.603627 | HPGD |
| 211549_s_at | 1.36E-05 | up | 2.012863 | HPGD |
| 238669_at | 2.01E-04 | up | 2.296154 | PTGS1 |
| 1554997_a_at | 9.37E-04 | up | 3.008369 | PTGS2 |
| 203913_s_at | 4.41E-09 | up | 4.03641 | HPGD |
| 203914_x_at | 2.93E-07 | up | 3.450221 | HPGD |
| 204748_at | 8.71E-06 | up | 7.748884 | PTGS2 |
| 236172_at | 1.25E-09 | up | 3.392918 | LTB4R |
| 204174_at | 1.11E-10 | up | 7.512037 | ALOX5AP |
| 204446_s_at | 2.35E-06 | up | 2.195747 | ALOX5 |
| 207206_s_at | 3.90E-05 | up | 2.604373 | ALOX12 |
| 214366_s_at | 2.90E-04 | up | 2.121638 | ALOX5 |
| 221027_s_at | 0.00175 | up | 2.000762 | PLA2G12A |
| 228084_at | 4.11E-04 | up | 2.14738 | PLA2G12A |
| 235394_at | 9.50E-06 | up | 2.173632 | PLAA |
| 242323_at | 5.76E-07 | up | 2.319156 | PLA2G12A |

Table 6. Cytokine

| Probe Set ID | Pvalue | Arrow | Fold | Gene |
|---|---|---|---|---|
| 237046_x_at | 4.92E-08 | down | -2.01547 | IL34 |
| 1568513_x_at | 5.20E-07 | down | -3.14638 | IL23A |
| 202859_x_at | 0.02858 | up | 2.698248 | IL8 |
| 203828_s_at | 2.70E-07 | down | -4.22368 | IL32 |
| 205067_at | 2.91E-09 | up | 11.1993 | IL1B |
| 206295_at | 2.08E-07 | up | 2.190391 | IL18 |
| 39402_at | 3.79E-08 | up | 9.375368 | IL1B |
| 203085_s_at | 2.46E-12 | down | -4.47642 | TGFB1 |
| 209651_at | 1.04E-05 | up | 3.862509 | TGFB1I1 |
| 205016_at | 3.08E-13 | up | 12.56279 | TGFA |
| 210354_at | 4.89E-05 | up | 2.295556 | IFNG |
| 230681_at | 3.27E-05 | down | -2.09638 | TBRG1 |

Table 7. Cytokine receptor

| Probe Set ID | Pvalue | Arrow | Fold | Gene |
|---|---|---|---|---|
| 239522_at | 1.63E-07 | down | -2.13821 | IL12RB1 |
| 201887_at | 2.63E-04 | up | 2.493641 | IL13RA1 |
| 201888_s_at | 3.50E-07 | up | 5.265817 | IL13RA1 |
| 202948_at | 3.01E-13 | up | 10.59575 | IL1R1 |
| 204116_at | 1.72E-13 | down | -4.5043 | IL2RG |
| 205403_at | 1.14E-14 | up | 37.11373 | IL1R2 |
| 205798_at | 0.001876 | down | -2.04842 | IL7R |
| 206618_at | 3.97E-05 | up | 2.889916 | IL18R1 |
| 209575_at | 6.50E-10 | up | 3.094987 | IL10RB |
| 210744_s_at | 5.37E-04 | up | 2.377707 | IL5RA |
| 210904_s_at | 6.93E-13 | up | 10.38528 | IL13RA1 |
| 211372_s_at | 8.57E-13 | up | 23.53884 | IL1R2 |
| 211517_s_at | 0.001136 | up | 2.419155 | IL5RA |
| 211612_s_at | 5.00E-09 | up | 8.944053 | IL13RA1 |
| 212196_at | 9.43E-05 | down | -2.08387 | IL6ST |
| 226333_at | 4.08E-13 | up | 2.730188 | IL6R |
| 224793_s_at | 4.59E-05 | up | 2.229928 | TGFBR1 |
| 225661_at | 1.28E-12 | up | 5.958052 | IFNAR1 |
| 225669_at | 8.99E-08 | up | 3.14582 | IFNAR1 |
| 204191_at | 2.17E-07 | up | 2.336802 | IFNAR1 |
| 242903_at | 1.66E-09 | down | -10.0312 | IFNGR1 |

Table 8. Complement

| Probe Set ID | Pvalue | Arrow | Fold | Gene |
|---|---|---|---|---|
| 200983_x_at | 2.01E-10 | up | 5.126428 | CD59 |
| 200984_s_at | 3.96E-10 | up | 5.39977 | CD59 |
| 200985_s_at | 7.88E-08 | up | 3.902797 | CD59 |
| 201925_s_at | 5.64E-10 | up | 4.165688 | CD55 |
| 201926_s_at | 7.26E-08 | up | 2.685217 | CD55 |
| 205382_s_at | 3.05E-04 | up | 3.283973 | CFD |
| 205654_at | 0.025737 | up | 2.728636 | C4BPA |
| 206244_at | 1.59E-08 | up | 5.148756 | CR1 |
| 208451_s_at | 1.33E-05 | up | 3.146097 | C4A/4B |
| 208488_s_at | 2.21E-06 | up | 2.838257 | CR1 |
| 208783_s_at | 4.44E-10 | up | 2.762472 | CD46 |
| 208910_s_at | 8.91E-07 | up | 2.772785 | C1QBP |
| 210184_at | 3.36E-07 | up | 2.689394 | ITGAX |
| 212463_at | 2.39E-08 | up | 2.387086 | CD59 |
| 217552_x_at | 1.26E-08 | up | 5.06804 | CR1 |
| 218232_at | 0.001617 | down | -2.31401 | C1QA |
| 218983_at | 5.79E-11 | up | 4.443039 | C1RL |
| 220088_at | 6.89E-05 | up | 4.115375 | C5AR1 |
| 239205_s_at | 5.31E-11 | up | 15.49001 | CR1/1L |
| 239206_at | 2.42E-09 | up | 12.87955 | CR1L |
| 244313_at | 8.98E-10 | up | 9.401143 | CR1 |
| 1555950_a_at | 8.18E-09 | up | 3.116447 | CD55 |

Table 9 anti-bacterial genes

| Probe Set ID | Pvalue | Arrow | Fold | Gene |
|---|---|---|---|---|
| 203591_s_at | 5.40E-05 | up | 2.953523 | CSF3R |
| 205118_at | 1.40E-06 | up | 6.013851 | FPR1 |
| 205119_s_at | 1.17E-07 | up | 8.646483 | FPR1 |
| 205159_at | 4.60E-08 | up | 10.71302 | CSF2RB |
| 206995_x_at | 9.73E-09 | up | 5.092197 | SCARF1 |
| 207085_x_at | 1.37E-06 | up | 3.863292 | CSF2RA |
| 207269_at | 9.38E-06 | up | 11.51901 | DEFA4 |
| 207677_s_at | 4.04E-08 | up | 5.465464 | NCF4 |
| 209949_at | 2.90E-05 | up | 2.761298 | NCF2 |
| 210340_s_at | 2.92E-06 | up | 3.513592 | CSF2RA |
| 210772_at | 3.39E-13 | up | 38.75074 | FPR2 |
| 210773_s_at | 1.67E-11 | up | 22.8151 | FPR2 |
| 211287_x_at | 5.85E-07 | up | 2.873763 | CSF2RA |
| 37408_at | 1.83E-05 | up | 2.611298 | MRC2 |
| 1552411_at | 1.45E-06 | up | 2.029896 | DEFB106A/B |
| 1553297_a_at | 1.38E-05 | up | 2.481309 | CSF3R |
| 206157_at | 9.63E-08 | up | 3.038519 | PTX3 |

Table 10 Glycolytic enzymes

| Probe Set ID | Pvalue | Arrow | Fold | Gene |
|---|---|---|---|---|
| 217356_s_at | 1.45E-07 | up | 3.485421 | PGK1 |
| 202464_s_at | 1.88E-05 | up | 3.848975 | PFKFB3 |
| 202934_at | 7.47E-04 | up | 2.054129 | HK2 |
| 202990_at | 4.01E-07 | up | 4.853299 | PYGL |
| 203502_at | 2.42E-08 | up | 19.10826 | BPGM |
| 206348_s_at | 4.04E-08 | up | 2.488083 | PDK3 |
| 209992_at | 1.30E-07 | up | 2.552616 | PFKFB2 |
| M33197_5_at | 4.35E-06 | up | 2.191367 | GAPDH |
| 217294_s_at | 9.10E-05 | up | 3.210724 | ENO1 |
| 217356_s_at | 1.45E-07 | up | 3.485421 | PGK1 |
| 213724_s_at | 2.15E-05 | up | 2.269032 | PDK2 |
| 225207_at | 0.04755 | up | 2.020549 | PDK4 |
| 227068_at | 5.94E-05 | up | 2.026827 | PGK1 |
| 228499_at | 1.41E-06 | up | 2.453812 | PFKFB4 |

Table11 ATPase

| Probe Set ID | Pvalue | Arrow | Fold | Gene |
|---|---|---|---|---|
| 200078_s_at | 1.87E-05 | up | 2.165182 | ATP6V0B |
| 200954_at | 7.36E-08 | up | 3.458313 | ATP6V0C |
| 201089_at | 2.00E-05 | up | 3.342557 | ATP6V1B2 |
| 201171_at | 2.74E-08 | up | 3.750572 | ATP6V0E1 |
| 201444_s_at | 8.20E-06 | up | 2.14297 | ATP6AP2 |
| 201971_s_at | 4.37E-04 | up | 2.028055 | ATP6V1A |
| 202872_at | 1.07E-10 | up | 6.460473 | ATP6V1C1 |
| 202874_s_at | 2.48E-10 | up | 4.5579 | ATP6V1C1 |
| 205704_s_at | 3.66E-05 | up | 2.185279 | ATP6V0A2 |
| 208898_at | 5.55E-05 | up | 2.130803 | ATP6V1D |
| 208899_x_at | 3.59E-08 | up | 2.716261 | ATP6V1D |
| 214149_s_at | 8.54E-11 | up | 7.664084 | ATP6V0E1 |
| 36994_at | 4.81E-09 | up | 2.599531 | ATP6V0C |
| 205950_s_at | 5.00E-08 | up | 42.28005 | CA1 |
| 206208_at | 2.02E-09 | up | 3.94974 | CA4 |
| 206209_s_at | 1.82E-12 | up | 8.938537 | CA4 |
| 209301_at | 2.90E-04 | up | 2.461168 | CA2 |

Table 12. Coagulation genes

| Probe Set ID | Pvalue | Arrow | Fold | Gene |
|---|---|---|---|---|
| 201108_s_at | 2.84E-05 | up | 3.303681 | THBS1 |
| 201109_s_at | 7.13E-05 | up | 2.95573 | THBS1 |
| 201110_s_at | 5.89E-08 | up | 6.730528 | THBS1 |
| 202112_at | 8.64E-07 | up | 2.154882 | VWF |
| 203887_s_at | 2.66E-10 | up | 13.1587 | THBD |
| 203888_at | 1.55E-09 | up | 5.29251 | THBD |
| 203989_x_at | 3.03E-05 | up | 3.24114 | F2R |
| 204713_s_at | 3.02E-06 | up | 2.665104 | F5 |
| 204714_s_at | 2.58E-06 | up | 3.393421 | F5 |
| 205756_s_at | 1.71E-07 | up | 2.887461 | F8 |
| 206429_at | 4.35E-06 | up | 2.168341 | F2RL1 |
| 206493_at | 0.012261 | up | 2.018667 | ITGA2B |
| 207808_s_at | 6.13E-06 | up | 3.830064 | PROS1 |
| 207815_at | 9.67E-06 | up | 8.676992 | PF4V1 |
| 207926_at | 1.53E-05 | up | 2.539035 | GP5 |
| 209769_s_at | 1.05E-08 | down | -2.30101 | GP1BB |
| 211661_x_at | 1.44E-06 | up | 5.09946 | PTAFR |
| 211924_s_at | 1.35E-07 | up | 4.396324 | PLAUR |
| 213258_at | 9.09E-06 | up | 3.442621 | TFPI |
| 213506_at | 1.11E-10 | up | 18.07023 | F2RL1 |
| 214866_at | 5.24E-08 | up | 3.066191 | PLAUR |
| 215240_at | 1.86E-07 | up | 4.259491 | ITGB3 |
| 219304_s_at | 0.001786 | up | 2.10823 | PDGFD |
| 219403_s_at | 0.002328 | up | 2.352545 | HPSE |
| 220336_s_at | 3.37E-04 | up | 2.506245 | GP6 |
| 222881_at | 1.96E-06 | up | 2.9733 | HPSE |
| 228618_at | 2.78E-05 | up | 2.611937 | PEAR1 |
| 231029_at | 2.14E-09 | up | 4.807453 | F5 |
| 236345_at | 1.63E-13 | up | 8.191023 | TBXAS1 |
| 237252_at | 1.11E-05 | up | 2.237747 | THBD |
| 242197_x_at | 6.23E-05 | down | -4.4414 | CD36 |

Table 13 RBC genes

| Probe Set ID | Pvalue | Arrow | Fold | Gene |
|---|---|---|---|---|
| 207793_s_at | 5.29E-06 | up | 3.527715 | EPB41 |
| 207854_at | 5.69E-04 | up | 2.700479 | GYPE |
| 208352_x_at | 5.02E-08 | up | 5.233521 | ANK1 |
| 208353_x_at | 1.46E-08 | up | 8.199239 | ANK1 |
| 208470_s_at | 4.92E-04 | up | 2.42323 | HP /// HPR |
| 209930_s_at | 7.44E-07 | up | 6.157681 | NFE2 |
| 211560_s_at | 5.08E-08 | up | 36.1193 | ALAS2 |
| 211820_x_at | 5.24E-10 | up | 17.19674 | GYPA |
| 211821_x_at | 5.49E-09 | up | 27.56794 | GYPA |
| 213515_x_at | 8.72E-07 | up | 13.31291 | HBG1 /// HBG2 |
| 215054_at | 4.51E-10 | up | 3.067622 | EPOR |
| 37986_at | 2.17E-09 | up | 2.737781 | EPOR |
| 215819_s_at | 1.12E-07 | up | 4.649047 | RHCE /// RHD |
| 216317_x_at | 0.001298 | up | 2.167189 | RHCE |
| 216563_at | 3.75E-05 | up | 2.423975 | ANKRD12 |
| 216833_x_at | 1.03E-06 | up | 7.30822 | GYPB /// GYPE |
| 218450_at | 1.29E-04 | up | 2.919147 | HEBP1 |
| 203665_at | 1.99E-04 | down | -2.47673 | HMOX1 |
| 219672_at | 1.22E-08 | up | 33.80057 | ERAF |
| 220807_at | 2.56E-10 | up | 18.14282 | HBQ1 |
| 223143_s_at | 0.001106 | up | 2.024285 | AKIRIN2 |
| 223542_at | 4.16E-06 | up | 2.408006 | ANKRD32 |
| 223669_at | 9.14E-08 | up | 20.2901 | HEMGN |
| 223670_s_at | 4.73E-11 | up | 49.78136 | HEMGN |
| 225051_at | 6.51E-05 | up | 2.198623 | EPB41 |
| 225735_at | 1.10E-06 | up | 2.178783 | ANKRD50 |
| 226663_at | 5.93E-06 | down | -2.50174 | ANKRD10 |
| 240336_at | 3.15E-12 | up | 61.68572 | HBM |
| 1554481_a_at | 4.42E-06 | up | 5.047096 | EPB41 |
| 204419_x_at | 3.33E-06 | up | 20.09402 | HBG1 /// HBG2 |
| 204848_x_at | 1.85E-06 | up | 17.25642 | HBG1 /// HBG2 |
| 206834_at | 0.011457 | up | 3.600573 | HBD |
| 213515_x_at | 8.72E-07 | up | 13.31291 | HBG1 /// HBG2 |